\documentclass[12pt]{iopart}
\usepackage[dvips]{graphicx}
\usepackage{latexsym}
\usepackage{iopams}
\usepackage{verbatim,times,bbm}
\usepackage{setstack}

\usepackage{color}

\begin{document}

\title{Capacities of lossy bosonic channel with correlated noise}
\author{Cosmo Lupo$^1$, Oleg V.~Pilyavets$^{1,2}$ and Stefano Mancini$^1$}
\address{$^1$Dipartimento di Fisica, Universit\`a di Camerino, I-62032 Camerino, Italy \\
$^2$P. N. Lebedev Physical Institute, Leninskii Prospect 53, Moscow
119991, Russia}

\eads{\mailto{cosmo.lupo@unicam.it}, \mailto{pilyavets@gmail.com},
\mailto{stefano.mancini@unicam.it}}

\begin{abstract}

We evaluate the information capacities of a lossy bosonic channel
with correlated noise. The model generalizes the one recently
discussed in [Phys.~Rev.~A {\bf 77}, 052324 (2008)], where memory
effects come from the interaction with correlated environments.
Environmental correlations are quantified by a multimode squeezing
parameter, which vanishes in the memoryless limit. We show that a
global encoding/decoding scheme, which involves input entangled
states among different channel uses, is always preferable with
respect to a local one in the presence of memory. Moreover, in a
certain range of the parameters, we provide an analytical expression
for the classical capacity of the channel showing that a global
encoding/decoding scheme allows to attain it. All the results can be
applied to a broad class of bosonic Gaussian channels.

\end{abstract}


\section{Introduction}\label{s:intro}

One of the main tasks of quantum information theory is the
evaluation of the capacities of quantum channels for the
transmission of classical or quantum information. Recently, a
growing attention has been devoted to the study of quantum channels
with memory. Coding theorems were provided for a subset of memory
channels \cite{werner}, the so-called `forgetful channels'. One can
distinguishes the cases in which the output at the $k$th use of the
channel is influenced by the input at the $k'$th use, with $k' < k$,
as the models studied in \cite{SD}; and those in which memory
effects come from correlations among subsequent channel uses, as the
ones considered in \cite{discr,cont,oleg,mancini}. Here we consider
the second case, which is also referred to as `channel with
correlated noise'. A correlation-free channel can be considered as
an ideal limit since correlations are unavoidable in physical
realizations. Another motivation for studying channels with
correlated noise is the possibility of enhancing the information
capacities. There are indeed evidences of the possibility of
amplifying the classical capacity in both the cases of discrete
\cite{discr} and continuous \cite{cont} variables quantum channels.

Here we consider a model of bosonic Gaussian channel in which memory
effects come from the interaction with a bosonic Gaussian
environment. The model is a generalization of the one discussed in
\cite{oleg} and it belongs to a family of channels presented in
\cite{mancini}. Even though each channel belonging to this family is
unitary equivalent to a memoryless one (in the sense specified in
\cite{mancini}), the presence of energy constraints can break the
unitary symmetry, leaving the problem of capacities evaluation open.
An instance of a channel belonging to that family is obtained by
specifying the state of the environment. Here we consider a
multimode squeezed thermal state, determined by two parameters. The
first parameter expresses the degree of squeezing, which in turn
determines the amount of correlations in the channel; the second one
is a temperature parameter expressing the mixedness of the state. It
is clear that at zero temperature the correlations in the multimode
squeezed state are quantum, on the other hand above a certain
temperature the states becomes separable and the correlations are
classical.

The choice of a Gaussian state for the environment makes in turns
the channel Gaussian. In this way, using
\cite{holevo,BroadBandC,Ent_ass_cap,Wolf}, we are able to evaluate,
analytically or numerically, the classical and quantum capacities of
the memory channel. To emphasize the role of correlations, we
compare two different scenarios for encoding and decoding classical
and quantum information: in the first one, which we refer to as the
{\it global scenario}, we allow preparation of states at the input
field which are entangled among different channel uses; in the
second one, called the {\it local scenario}, we only allow
preparation of simply separable states (i.e.\ uncorrelated) at the
input field, moreover we do not allow the receiver to access the
correlations among the output modes.

The paper develops along the following lines. In section \ref{model}
we present the model and define the global and the local
encoding/decoding scenarios. In section \ref{capacities} we present
analytical and numerical results for the classical,
entanglement-assisted, and quantum capacity. Conclusions and
comments are drawn in section \ref{conclude}.

\section{A model of lossy bosonic Gaussian channel with correlated noise}\label{model}

\begin{figure}
\centering
\includegraphics[width=0.4\textwidth]{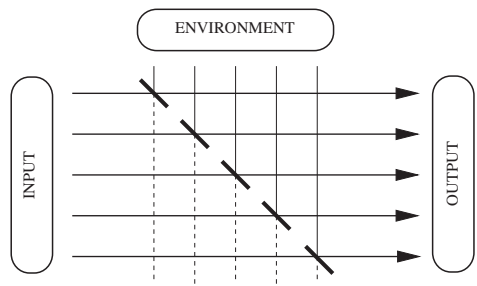}
\caption{A schematic picture of the model of lossy bosonic channel.
Each input mode (left-right line), representing one use of the
channel, interacts with the corresponding environment mode
(top-bottom line) through a beam-splitter. To introduce correlations
effects, environment modes are considered in a correlated state.}
\label{fig_model}
\end{figure}

We consider an instance of the general model for a bosonic channel
with correlated noise presented in \cite{mancini}. For any integer
$n$, its action is defined over a set of $n$ input bosonic
oscillators, with canonical variables $\{ q_k, p_k \}_{k=1, \dots
n}$. A collection of ancillary modes $\{ Q_k, P_k \}_{k=1, \dots
n}$, which play the role of the environment, is also needed. In the
following we refer to this set of oscillators as the input and
environment {\em local modes}. All the frequencies are assumed to be
degenerate and equal to one, together with $\hbar=1$. The integer
$k$ labels the sequential uses of the channel. At the $k$th use, the
$k$th input mode is linearly mixed with the $k$th mode of the
environment at a beam splitter with given transmissivity $\eta$ (see
figure \ref{fig_model}). In the Heisenberg picture the channel
transforms the input field variables as
\begin{eqnarray}
\begin{array}{ccc}
q'_k & = & \sqrt{\eta} \, q_k + \sqrt{1-\eta} \, Q_k \, ,\\
p'_k & = & \sqrt{\eta} \, p_k + \sqrt{1-\eta} \, P_k \, .
\end{array}
\end{eqnarray}
A constraint on the energy is required to avoid infinite capacities.
We constraint the average number of photons at the input field; for
a given $N$ we require:
\begin{equation}\label{energy}
\frac{1}{2n} \sum_{k=1}^n \langle q_k^2 + p_k^2 \rangle_\mathrm{in}
\le N + \frac{1}{2}.
\end{equation}

Memory effects appear in the presence of correlations among the
local modes of the environment. For a given integer $n$, a channel
$\mathfrak{L}^{(n)}$ over the $n$ input modes is defined. In the
Schroedinger picture its action is
\begin{equation}
\mathfrak{L}^{(n)}(\rho_\mathrm{in}) = \mathrm{tr}_\mathrm{env}
\left( \mathcal{U} \ \rho_\mathrm{in} \hspace{-1mm} \otimes
\hspace{-0.5mm} \rho_\mathrm{env} \ \mathcal{U}^\dag \right)
\end{equation}
where $\rho_\mathrm{in}$ indicates the state of the input field,
$\mathcal{U}$ is the unitary transformation at the $n$ beam
splitters, $\rho_\mathrm{env}$ indicates the state of the
environment field and $\mathrm{tr}_\mathrm{env}$ the partial trace
over the environment variables. The channel is correlation-free if
the state of the environment is simply separable (i.e.\
uncorrelated) in the basis of the local mode.

We assume the environment to be in a Gaussian state, which in turns
makes the channel Gaussian. Here we consider an environment
covariance matrix of the following {\em block-diagonal} form:
\begin{eqnarray}\label{covm}
V = \left(
\begin{array}{cc}
\langle \mathbf{Q} \mathbf{Q}^\mathsf{T} \rangle & \langle \frac{\mathbf{Q}\mathbf{P}^\mathsf{T}+\mathbf{P}\mathbf{Q}^\mathsf{T}}{2} \rangle \\
\langle
\frac{\mathbf{Q}\mathbf{P}^\mathsf{T}+\mathbf{P}\mathbf{Q}^\mathsf{T}}{2}
\rangle & \langle \mathbf{P} \mathbf{P}^\mathsf{T} \rangle
\end{array}
\right) := \left(T+\frac{1}{2}\right) \left(
\begin{array}{cc}
e^{s\Omega} & \mathbb{O} \\
\mathbb{O}  & e^{-s\Omega}
\end{array} \right),
\end{eqnarray}
where $\mathbf{Q} := (Q_1, Q_2, \dots Q_n)^\mathsf{T}$ and
$\mathbf{P} := (P_1, P_2, \dots P_n)^\mathsf{T}$. This is a {\it
bona fide} covariance matrix as long as the matrix $\Omega$ is
symmetric and $T \ge 0$. To fix the ideas we chose a $n \times n$
matrix $\Omega$ of the following form:
\begin{eqnarray}\label{omega}
\Omega = \left( \begin{array}{cccccc}
0      & 1      & 0      & \dots  & 0      & 0 \\
1      & 0      & 1      & \ddots & 0      & 0 \\
0      & 1      & 0      & \ddots & 0      & 0 \\
\vdots & \vdots & \ddots & \ddots & \vdots & \vdots \\
0      & 0      & 0      & \ddots & 0      & 1 \\
0      & 0      & 0      & \dots  & 1      & 0
\end{array} \right).
\end{eqnarray}
In this way, for $T=0$, we recover the model analyzed in
\cite{oleg}. The physical interpretation is that of a multimode
squeezed thermal state, that can be obtained by applying the
multimode squeezing operator (see \cite{squeez})
\begin{equation}\label{u_squeeze}
U(s) = \exp{\left[- i \frac{s}{2} \sum_{k=1}^{n-1} \left( Q_k
P_{k+1} + P_k Q_{k+1} \right) \right]}
\end{equation}
to a thermal state with $T$ average excitations per mode. Since the
memory of the channel comes from squeezing in the environment we
refer to the parameter $s$ (or $|s|$) as the {\it memory parameter};
for $s=0$ the environment is a correlation-free thermal state. The
integer $n$ can be interpreted as the characteristic length of
correlations. To evaluate lower bounds we need the eigenvalues and
eigenvectors of the matrix $\Omega$ which were presented in
\cite{oleg}. The eigenvalues are
\begin{equation}
\lambda_j = 2 \cos{\left( \frac{\pi j}{n+1}\right)} \ \ \mbox{for} \
\ j=1, \dots n,
\end{equation}
where the corresponding eigenvectors have components
\begin{equation}\label{evectors}
v_{j,k} = \sqrt{\frac{2}{n+1}} \sin{\left(\frac{jk\pi}{n+1}\right)}
\ \ \mbox{for} \ \ k=1,\dots n.
\end{equation}

Let us introduce the global variables $\{\tilde{Q}_j, \tilde{P}_j
\}_{j=1,\dots n}$, defined from the eigenvectors of $\Omega$ as:
\begin{eqnarray}\label{collective}
\begin{array}{ccc}
\tilde{Q}_j & := & \sum_{k} v_{j,k} \, Q_k \, ,\\
\tilde{P}_j & := & \sum_{k} v_{j,k} \, P_k \, .
\end{array}
\end{eqnarray}
It follows that the environment covariance matrix is diagonal in
this basis; it can be written as a direct sum
\begin{equation}
\tilde{V} = \bigoplus_{j=1}^n \tilde{V}_j
\end{equation}
of single-mode covariance matrices of the form:
\begin{eqnarray}\label{coll_modek}
\tilde{V}_j = \left(
\begin{array}{cc}
\langle \tilde{Q}_j^2 \rangle & \langle \frac{\tilde{Q}_j\tilde{P}_j+\tilde{P}_j\tilde{Q}_j}{2} \rangle \\
\langle \frac{\tilde{Q}_j\tilde{P}_j+\tilde{P}_j\tilde{Q}_j}{2}
\rangle & \langle \tilde{P}_j^2 \rangle
\end{array}
\right) = \left( T + \frac{1}{2}\right) \left( \begin{array}{cc} e^{s_j} & 0 \\
0 & e^{-s_j}
\end{array} \right),
\end{eqnarray}
with $s_j := s \lambda_j$. Hence, moving to the global variables,
the state of the environment is the direct product of $n$ modes,
each being in a squeezed thermal state, with squeezing parameter
$s_j$ and $T$ thermal photons.

\subsection{Global versus local scenario}\label{glocal}

In the following sections we estimate the quantum and classical
capacities for a given value of the correlation length $n$. To
emphasize the role of correlations we compare two different
encoding/decoding scenarios.

The first scenario is a global one, in the sense that it involves
preparation of the input field in states which are (in general)
entangled among different uses of the channel (i.e.\ among the local
modes). The second scenario is a local one, involving
preparation of the input field in states which are simply separable
among the channel uses; moreover we do not allow the receiver to
access the correlations among different local modes at the output
field.

As to the global scenario, we introduce the following global
variables at the input field
\begin{eqnarray}\label{point}
\begin{array}{ccc}
\tilde{q}_j & := & \sum_{k} v_{j,k} \, q_k \, ,\\
\tilde{p}_j & := & \sum_{k} v_{j,k} \, p_k \, ,
\end{array}
\end{eqnarray}
(notice that this set of variables is `parallel' to the environment
global variables defined in (\ref{collective})) and consider states
of the input field which are factorized in the basis of these global
modes. These states are in general entangled among channel uses
(i.e.\ among local modes). In terms of the global variables, the
channel factorizes as
\begin{equation}
\mathfrak{L}^{(n)} = \bigotimes_j \tilde{\mathfrak{L}}^{(1)}_j \, ,
\end{equation}
where the channel $\tilde{\mathfrak{L}}^{(1)}_j$ acts on the $j$th
global mode of the input field. In the Heisenberg picture
$\tilde{\mathfrak{L}}^{(1)}_j$ transforms the $j$th global input
variables as:
\begin{eqnarray}
\begin{array}{ccc}
\tilde{q}'_j & = & \sqrt{\eta} \, \tilde{q}_j + \sqrt{1-\eta} \, \tilde{Q}_j \, ,\\
\tilde{p}'_j & = & \sqrt{\eta} \, \tilde{p}_j + \sqrt{1-\eta} \,
\tilde{P}_j \, .
\end{array}
\end{eqnarray}
Furthermore, due to the form of the transformation (\ref{point}) (a
linear passive one, see \ref{appxA}), the energy constraint is
preserved in the basis of global variables:
\begin{equation}
\frac{1}{2n} \sum_{j=1}^n \langle \tilde{q}_j^2 + \tilde{p}_j^2
\rangle_\mathrm{in} \le N + \frac{1}{2}.
\end{equation}
It is worth noticing that the channel $\tilde{\mathfrak{L}}^{(1)}_j$
is a lossy bosonic channel in which one (global) mode is mixed with
an environment mode which is squeezed (it is described by the
covariance matrix (\ref{coll_modek})). Hence the $n$-use channel is
unitary equivalent to a correlation-free channel, which is the
product of $n$ single-mode channels.

Concerning the local scenario let us say that, as long as the
correlations among the local modes at the output field are
neglected, the environment can be effectively described by a state
which is factorized in the local modes. For each $k$, by
integrating the environment Wigner function over the local variables
$\{ Q_h , P_h \}$ for $h \neq k$, we obtain the Wigner function of
the $k$th local mode of the environment. The corresponding state is
thermal-like, the average number of photon can be computed from
(\ref{collective}), yielding:
\begin{equation}
T_\mathrm{eff}(k) = \left(T+\frac{1}{2}\right) \left[ \sum_{j=1}^n
|v_{j,k}|^2 e^{s_j} \right] - \frac{1}{2}.
\end{equation}
By the symmetries of $v_{j,k}$ and $s_j$, this can be rewritten as
follows:
\begin{eqnarray}\fl
T_\mathrm{eff}(k) = \left\{
\begin{array}{lr}
(2T+1)\left(\sum_{j=1}^{n/2} |v_{j,k}|^2 \cosh{s_j}\right) - \frac{1}{2} & \mbox{if $n$ is even,}\\
(2T+1)\left(\sum_{j=1}^{(n-1)/2} |v_{j,k}|^2 \cosh{s_j}\right) +
\frac{1}{2} |v_{(n+1)/2,k}|^2  - \frac{1}{2} & \mbox{if $n$ is odd.}
\end{array}
\right.
\end{eqnarray}
Hence in the local scenario the state of the environment can be
substituted with a thermal state with $k$-dependent temperature.
Notice that, as one can expect, the {\em local temperature}
$T_\mathrm{eff}(k)$ monotonically increases with $|s|$.

\section{Evaluating capacities}\label{capacities}

This section is devoted to the evaluation of classical and quantum
capacities of the channel $\mathfrak{L}^{(n)}$. This requires the
constrained optimization of several entropic quantities, as the
Holevo information, the quantum mutual information, the coherent
information.

Let us recall that the von Neumann entropy $S(\rho)$ of a $n$ mode
Gaussian state $\rho$ can be computed as follows:
\begin{equation}
S(\rho) = \sum_{a=1}^n g( \nu_a - 1/2 )
\end{equation}
where
\begin{equation}
g(x) := (x+1)\log_2{(x+1)} - x \log_2{x}
\end{equation}
and $\nu_a$ are the $n$ symplectic invariants, i.e.\ the symplectic
eigenvalues of the covariance matrix (see e.g.\ \cite{paris}). In
the case of a single mode with covariance matrix $\sigma$ the only
symplectic invariant is $\det(\sigma)$, and the symplectic
eigenvalue is $\nu = \sqrt{\det(\sigma)}$. It is worth remarking
that both in the global and local scenarios introduced above the
evaluation of the entropy of the $n$-mode fields reduces to the
single-mode case.

As to the evaluation of the classical capacity of a quantum channel
$\mathfrak{L}$, one is led to the Holevo information, defined as
\begin{equation}
\chi(\mathfrak{L},\{\rho_\alpha,dp_\alpha\}) :=
S(\mathfrak{L}(\rho)) - \int dp_\alpha S(\mathfrak{L}(\rho_\alpha)),
\end{equation}
where $\rho_\alpha$ denotes the quantum state encoding the classical
variable $\alpha$, with probability density $dp_\alpha$, and $\rho$
is the ensemble state, $\rho=\int dp_\alpha \rho_\alpha$.

The evaluation of quantum capacity and entanglement assisted
classical capacity involves the coherent information, defined as
\begin{equation}
J(\mathfrak{L},\rho) := S(\mathfrak{L}(\rho)) -
S(\mathfrak{L},\rho).
\end{equation}
The quantity denoted $S(\mathfrak{L},\rho)$ is the entropy exchange,
defined as
\begin{equation}
S(\mathfrak{L},\rho) :=
S(\mathfrak{L}\otimes\mathfrak{I}(\tilde{\rho})),
\end{equation}
where $\tilde{\rho}$ is a purification of the state $\rho$ involving
an ancillary system, and $\mathfrak{I}$ is the identical channel
acting on the ancilla.

It is worth remarking that Gaussian encoding it is known to be
optimal for classical \cite{BroadBandC,Ent_ass_cap} and quantum
capacities \cite{Wolf} in the memoryless case. Motivated by these
results we are going to estimate the capacities of the memory
channel using Gaussian encoding.

Concerning the global scenario, the maximization of the entropic
functions is performed over a set of Gaussian states which are
simply separable with respect to the global modes. Notice that these
states are in general entangled in the local basis, i.e.\ entangled
among channel uses. In particular, we consider covariance matrices
of the form
\begin{equation}\label{in_cov}
\sigma = \bigoplus_{j=1}^n \sigma_j,
\end{equation}
where $\sigma_j$ is the covariance matrix of the $j$th global input
mode. For real $r_j$ and $t_j \ge 0$, the $j$th covariance matrix is
chosen as follows:
\begin{eqnarray}\label{in_cov_par}
\sigma_j = \left(
\begin{array}{cc}
\langle \tilde{q}_j^2 \rangle & \langle \frac{\tilde{q}_j\tilde{p}_j+\tilde{p}_j\tilde{q}_j}{2} \rangle \\
\langle \frac{\tilde{q}_j\tilde{p}_j+\tilde{p}_j\tilde{q}_j}{2}
\rangle & \langle \tilde{p}_j^2 \rangle
\end{array}
\right) := \left( t_j + \frac{1}{2} \right) \left(
\begin{array}{cc}
e^{r_j} & 0       \\
0       & e^{-r_j}
\end{array}
\right).
\end{eqnarray}
Under the action of the channel $\tilde{\mathfrak{L}}^{(1)}_j$ this
matrix is mapped into the covariance matrix
\begin{equation}
\sigma_j' = \eta \sigma_j + (1-\eta) V_j.
\end{equation}

The energy constraint can be written in terms of the input
covariance matrix as
\begin{equation}
\frac{1}{2n} \sum_{j=1}^n \mathrm{tr}\left( \sigma_j \right) \le N +
\frac{1}{2}
\end{equation}
that is
\begin{equation}\label{E_const}
\frac{1}{n} \sum_{j=1}^n \left( t_j + \frac{1}{2} \right) \cosh{r_j}
\le N + \frac{1}{2}.
\end{equation}
It can be useful to write the energy constraints in two steps,
namely
\begin{eqnarray}
\left( t_j + \frac{1}{2} \right) \cosh{r_j} = N_j + \frac{1}{2} \\
\frac{1}{n} \sum_{j=1}^n N_j \le N.
\end{eqnarray}

Concerning the local scenario, as noticed above, the channel reduces
to a correlation-free channel with thermal environment, with a
$k$-dependent effective temperature. For this kind of channel,
expressions for the (one-shot) classical and quantum capacities are
available in literature \cite{holevo,Wolf} and will be used as a
term of comparison.

\subsection{Holevo information}\label{ss:cc}

First, let us compute an upper bound for the classical capacity of
the memory channel. This can be obtained by the maximal output
entropy. For given $n$, we have
\begin{equation}
C \le \frac{1}{n} \sup S(\mathfrak{L}^{(n)}(\rho)) \le \frac{1}{n}
\sum_{j=1}^n \sup S(\tilde{\mathfrak{L}}_j^{(1)}(\rho_j)) =: C^>,
\end{equation}
where the superior is over all input states $\rho$ ($\rho_j$)
satisfying the energy constraints, which is reached by Gaussian
input states. The maximum output entropy is reached by input states
as in (\ref{in_cov_par}). The contribution of the $j$th global mode
to the output entropy reads
\begin{equation}
S(\tilde{\mathfrak{L}}_j^{(1)}(\rho_j)) = g\left(
\sqrt{\det{(\sigma_j')}}-1/2 \right),
\end{equation}
where
\begin{eqnarray}
\det{(\sigma_j')} & = & \left( \eta (t_j+1/2) e^{r_j} + (1-\eta)
(T+1/2) e^{s_j} \right)
\nonumber\\
& \times & \left( \eta (t_j+1/2) e^{-r_j} + (1-\eta) (T+1/2)
e^{-s_j} \right). \nonumber
\end{eqnarray}
For sufficiently small values of $s$, it is possible to write an
explicit solution. The maximum output entropy is reached in
correspondence of the optimal values satisfying
\begin{equation}
\left( t_j^\mathrm{opt} + \frac{1}{2} \right) e^{\pm
r_j^\mathrm{opt}} = N_j^\mathrm{opt} + \frac{1}{2} \mp
\frac{1-\eta}{\eta}\left( T + \frac{1}{2} \right) \sinh{s_j},
\end{equation}
with
\begin{equation}\label{N_opt}
N_j^\mathrm{opt} = N - \frac{1-\eta}{\eta}\left(T+\frac{1}{2}\right)
\cosh{s_j} + \frac{1-\eta}{\eta} \left(M(s,T)+\frac{1}{2}\right),
\end{equation}
and
\begin{equation}\label{M_quantity}
M(s,T) := \frac{1}{n} \left(T+\frac{1}{2}\right) \left( \sum_{k=1}^n
\cosh{(s_k)} \right) - \frac{1}{2}.
\end{equation}
The range in which these values are optimal is determined by the
relations $t_j^\mathrm{opt}\ge 0$, namely:
\begin{equation}\label{upper_range}
\left(N_j^\mathrm{opt}+\frac{1}{2}\right)^2 -
\left(\frac{1-\eta}{\eta}\right)^2 \left(T+\frac{1}{2}\right)^2
\sinh^2{s_j} \ge \frac{1}{4}.
\end{equation}
In this range the upper bound can be computed analytically, yielding
\begin{equation}\label{upper}
C^> = g\left[ \eta N + (1-\eta) M(s,T) \right].
\end{equation}

Let us now compute a lower bound for the classical capacity.
Motivated by the fact that it is optimal for the memoryless channel
\cite{BroadBandC} we consider encoding in displaced states:
\begin{equation}\label{seed}
\varrho_\alpha = \mathcal{D}(\alpha) \varrho
\mathcal{D}^\dag(\alpha).
\end{equation}
Here $\varrho$ denotes a Gaussian seed state, not necessarily the
vacuum, of the $n$ mode input field. As to the global scenario, its
covariance matrix is assumed to be as in equations (\ref{in_cov}),
(\ref{in_cov_par}). The displacement operator can be decomposed in
the basis of global modes as
\begin{equation}
\mathcal{D}(\alpha) = \bigoplus_{j=1}^n \mathcal{D}_j(\alpha_j),
\end{equation}
where $\alpha_j = (y_{q,j} + i y_{p,j})/\sqrt{2}$ is the
displacement amplitude of the $j$th global input mode, and $\alpha
:= (\alpha_1, \dots \alpha_n)$. The classical noise is assumed to be
Gaussian with zero mean and covariance matrix $Y$ of the following
form:
\begin{equation}
Y = \bigoplus_{j=1}^n Y_j \, ,
\end{equation}
where, for $c_{q,j}, c_{p,j} \geq 0$,
\begin{eqnarray}\label{sigma}
Y_j = \left(
\begin{array}{cc}
\langle y_{q,j}^2 \rangle & \langle y_{q,j} y_{p,j} \rangle \\
\langle y_{q,j} y_{p,j} \rangle       & \langle y_{p,j}^2 \rangle
\end{array}
\right) := \left(
\begin{array}{cc}
c_{q,j} & 0 \\
0       & c_{p,j}
\end{array}
\right).
\end{eqnarray}

At the $j$th global mode the classical noise induces an ensemble
state described by the covariance matrix
\begin{equation}
\bar{\sigma}_j := \sigma_j + Y_j,
\end{equation}
that, under the action of the channel
$\tilde{\mathfrak{L}}^{(1)}_j$, is mapped into
\begin{equation}
\bar{\sigma}_j' = \eta \bar{\sigma}_j + (1-\eta) V_j.
\end{equation}

In our setting, the lower bound for the classical capacity is
computed by maximizing the Holevo information
\begin{equation}\label{chi}
\chi = \sum_{j=1}^n \chi_j = \sum_{j=1}^n
g\left(\sqrt{\det(\bar{\sigma}_j')} - 1/2\right) -
g\left(\sqrt{\det(\sigma_j')} - 1/2\right)
\end{equation}
over the $4n$ parameters $\{ t_j, r_j, c_{q,j}, c_{p,j}\}$ under the
energy constraints
\begin{eqnarray}
\frac{1}{2} \left( c_{q,j} + c_{p,j} \right) + \left( t_j + 1/2 \right) \cosh{r_j} = N_j + \frac{1}{2} \\
\frac{1}{n} \sum_{j=1}^n N_j \le N.
\end{eqnarray}
For given values of $N$, $\eta \in (0,1)$, $T$ and for sufficiently
small values of $s$ we can write an explicit solution. The maximum
Holevo information is reached in correspondence of the following
optimal values of the parameters: $t_j^\mathrm{opt}=0$,
$r_j^\mathrm{opt}=s_j$,
\begin{equation}
c_{q,j}^\mathrm{opt} = N_j^\mathrm{opt}+\frac{1}{2} -
\frac{e^{s_j}}{2} -
\frac{1-\eta}{\eta}\left(T+\frac{1}{2}\right)\sinh{s_j}
\end{equation}
and
\begin{equation}
c_{p,j}^\mathrm{opt} = N_j^\mathrm{opt}+\frac{1}{2} -
\frac{e^{-s_j}}{2} +
\frac{1-\eta}{\eta}\left(T+\frac{1}{2}\right)\sinh{s_j},
\end{equation}
where $N_j^\mathrm{opt}$ is as in equation (\ref{N_opt}).

\begin{figure}
\centering
\includegraphics[height=0.4\textwidth]{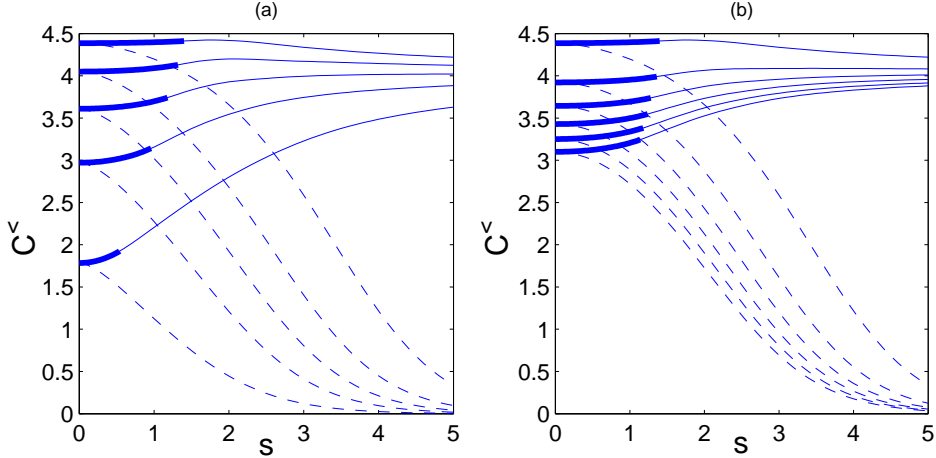}
\caption{The plots show the lower bounds for the classical capacity,
for $n=10$, as function of the memory parameter $|s|$. In (a) at
$T=0$, where the lower bound is strict, for different values of
$\eta$, from bottom to top $\eta$ varies from $0.1$ to $0.9$ by
steps of $0.2$. In (b) at $\eta=0.9$ for different values of $T$,
from top to bottom $T$ varies from $0$ to $5$ by steps of $1$. The
solid lines refer to the global scenario, the tick ones refer to the
analytical solution in the region in which it is available. The
lower bounds for the local scenario are plotted in dashed lines. The
maximum average number of excitations per mode in the input field is
$N=8$.} \label{Classical}
\end{figure}

The range of $s$ for which these values are optimal is defined by
the conditions $c_{q,j}^\mathrm{opt} \ge 0$, $c_{p,j}^\mathrm{opt}
\ge 0$. In that range, we are able to provide the following
analytical lower bound for the classical capacity per channel use:
\begin{equation}\label{class}
C^< = g\left[ \eta N + (1-\eta) M(s,T) \right] - g\left[ (1-\eta) T
\right].
\end{equation}
For $T=0$, this improves the lower bound computed in \cite{oleg}.
From the lower bound we can deduce that, besides the trivial cases
$\eta(1-\eta) = 0$, the classical capacity monotonically increases
with $|s|$. Moreover, it is worth noticing that, at $T=0$, the lower
bound (\ref{class}) coincides with the upper bound (\ref{upper}).
Hence, in the intersection of their ranges of validity the
analytical upper and lower bounds are strict and the expression in
(\ref{class}) is the capacity of the memory channel at zero
environment temperature. It is easy to see that the range of
validity of the upper bound (\ref{upper_range}) contains the one of
the lower bound, thus the lower bound in (\ref{class}) is strict in
the whole range of its validity at $T=0$. Notice that the results
can be extended to the limit $n\rightarrow\infty$ as in \cite{oleg}.
For higher values of $|s|$ one can look for a numerical solution.
Figure \ref{Classical} shows the analytical and numerical lower
bound as function of the memory parameter for several values of
$\eta$ and $T$.

It is interesting to consider the limit $|s| \rightarrow \infty$
corresponding to `infinite correlations'. Let us consider the term
$\chi_j$ in (\ref{chi}) coming from the contribution of the $j$th
global mode. Without loss of generality, we can assume $s_j
>0$. In the limit $s_j \gg 1$ we can write the following asymptotic
expressions:
\begin{eqnarray}
\det(\sigma_j') & \simeq & \eta(1-\eta)\left(T+\frac{1}{2}\right)\left(t_j+\frac{1}{2}\right)e^{s_j-r_j} + O(e^{-s_j}), \label{vout}\\
\det(\bar{\sigma}_j') & \simeq &
\eta(1-\eta)\left(T+\frac{1}{2}\right)\left[\left(t_j+\frac{1}{2}\right)e^{-r_j}+
c_{p,j} \right] e^{s_j} +  O(e^{-s_j}). \label{voutave}
\end{eqnarray}
By noticing that
\begin{equation}
\lim_{x\rightarrow\infty} \left[ g(x) -
\left(\log_2{x}-\log_2{e}\right) \right]=0,
\end{equation}
we obtain the following asymptotic expression for the $j$th term in
the Holevo information:
\begin{equation}\label{mutual_infty}
{\chi_\infty}_j = \lim_{s\rightarrow\infty} \chi_j = \frac{1}{2}
\log_2{\left(1 + \frac{c_{p,j} e^{r_j}}{t_j+1/2}\right)}.
\end{equation}
For given value of $N_j$, from the last expression one obtains that
the maximum of the mutual information is reached for
$t^\mathrm{opt}_j=0$, $r^\mathrm{opt}_j=\ln (2N_j+1)$,
$c^\mathrm{opt}_{q,j}=0$, and $c^\mathrm{opt}_{p,j} =
\sinh{r^\mathrm{opt}_j}$, yielding the following value for the $j$th
contribution to the classical capacity:
\begin{equation}
{C^<_\infty}_{j} = \max_{\{t_j, r_j, c_{q,j}, c_{p,j}\}}
\chi_j^\infty = \log_2{(2N_j+1)}.
\end{equation}
Summing over $j$ we obtain the following expression for the capacity
per channel use:
\begin{eqnarray}\label{class_infty}\fl
C^<_\infty = \left\{ \begin{array}{lc}
\log_2{(2N+1)} & \mbox{if $n$ is even,} \\
\frac{n-1}{n} \log_2{(2N+1)} + \frac{1}{n}\left\{g\left[ \eta N +
(1-\eta) T \right] - g\left[ (1-\eta) T \right]\right\} & \mbox{if
$n$ is odd.}
\end{array}\right.
\end{eqnarray}
The presence of an extra term for odd $n$ comes from the
contribution of the $s_j=0$ term and it leads to the oscillations of
the Holevo information with the number of uses already observed in
\cite{oleg}. However, the relative amplitude of these oscillations
becomes negligible as the number of uses increases. Interestingly
enough, in the limit of perfect memory the maximal Holevo
information is determined solely by the value of $N$, i.e.\ by the
energy constraints. The asymptotic lower bound can be reached by
homodyne detection (see \cite{ICQO}).

To conclude this section, let us mention that a lower bound
concerning the local scenario can be obtained by the following
expression (see \cite{mancini}):
\begin{equation}\nonumber\fl
C^< = \frac{1}{n} \max_{\{ N_k \}} \left\{ \sum_{k=1}^n g[\eta N_k +
(1-\eta)T_\mathrm{eff}(k)] - g[(1-\eta)T_\mathrm{eff}(k)] \ \ | \ \
\frac{1}{n} \sum_{k=1}^n N_k = N \right\}.
\end{equation}
This lower bound saturates the channel capacity for
$T_\mathrm{eff}=0$, see \cite{BroadBandC}, which is obtained for
$s=0$, $T=0$. This bound is plotted in figure \ref{Classical}
together with the lower bound computed in the global scenario.

\subsection{Coherent information, quantum mutual
information}\label{ss:qc}

The problem of evaluating the quantum capacity is greatly simplified
by the fact that the channel $\mathfrak{L}^{(n)}$ is degradable for
$\eta \in [1/2,1]$ and anti-degradable for $\eta\in[0,1/2)$. It
follows that the coherent information is additive for $\eta \in
[1/2,1]$ and the quantum capacity vanishes for $\eta\in[0,1/2[$ (see
\cite{degrad,Caruso,Wolf}). It is easy to recognize that this
property is shared by all the Gaussian memory channels of the kind
presented in \cite{mancini}. For the same reason, in the global
scenario, the $n$-mode channel reduces to the single-mode case as
\begin{equation}
\sup_\rho J(\mathfrak{L}^{(n)},\rho) = \sup_{\{\rho_j\}}
\sum_{j=1}^n J(\tilde{\mathfrak{L}}^{(1)}_j,\rho_j),
\end{equation}
with the proper energy constraint.

\begin{figure}
\centering
\includegraphics[height=0.4\textwidth]{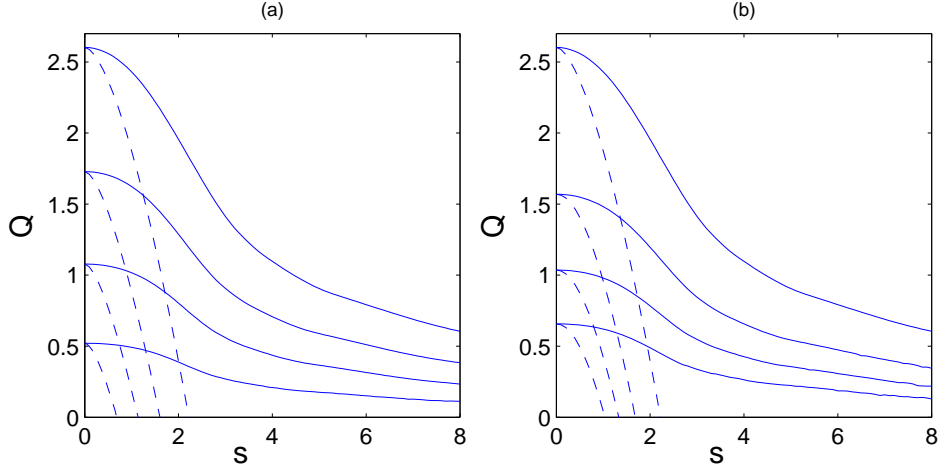}
\caption{The plots show the numerically evaluated quantum capacity,
for $n=10$, as function of the memory parameter $|s|$. In (a) at
$T=0$ for different values of $\eta$, from bottom to top $\eta$
varies from $0.6$ to $0.9$ by steps of $0.1$. In (b) at $\eta=0.9$
for different values of $T$, from top to bottom $T$ varies from $0$
to $1.5$ by steps of $0.5$. The solid lines refer to the global
scenario, the dashed lines to the local one. The maximum average
number of excitations per mode in the input field is $N=8$.}
\label{Quantum}
\end{figure}

For the quantum capacity, and for the entanglement-assisted
classical capacity, we need to evaluate the entropy exchange of the
channel $\tilde{\mathfrak{L}}_j^{(1)}$. It follows from \cite{Wolf},
and from \cite{holevo,Ent_ass_cap} that it is sufficient to consider
Gaussian input states. Numerical analysis shows that the choice of
input state with covariance matrix of the form (\ref{in_cov_par}) is
optimal. The input state at the $j$th global mode, with covariance
matrix as in (\ref{in_cov_par}), can be purified into a two mode
Gaussian state with covariance matrix:
\begin{eqnarray}
\tau_j = \left(
\begin{array}{cccc}
a_j & 0    & x_j & 0    \\
0   & b_j  & 0   & -x_j \\
x_j & 0    & b_j & 0    \\
0   & -x_j & 0   & a_j
\end{array}
\right)
\end{eqnarray}
where
\begin{equation}
a_j := (t_j+1/2)e^{r_j}, \quad b_j := (t_j+1/2)e^{-r_j}, \quad x_j
:= \sqrt{a_j b_j - 1/4}.
\end{equation}
The action of the channel $\tilde{\mathfrak{L}}_j^{(1)} \otimes
\mathfrak{I}$ leads to the output covariance matrix:
\begin{eqnarray}\label{pure_out}\fl
\tau'_j = \left(
\begin{array}{cc}
A_j & C_j^\mathsf{T} \\ C_j & B_j
\end{array}
\right) := \left(
\begin{array}{cccc}
\eta a_j + (1-\eta)c_j & 0                      & \sqrt{\eta} x_j & 0                \\
0                      & \eta b_j + (1-\eta)d_j & 0               & -\sqrt{\eta} x_j \\
\sqrt{\eta} x_j        & 0                      & b_j             & 0                \\
0                      & -\sqrt{\eta} x_j       & 0               &
a_j
\end{array}
\right)
\end{eqnarray}
where
\begin{equation}
c_j := (T+1/2)e^{s_j}, \quad d_j = (T+1/2)e^{-s_j}.
\end{equation}
The symplectic eigenvalues of the covariance matrix in
(\ref{pure_out}) are:
\begin{equation}
\nu_{j,\pm} = \frac{1}{\sqrt{2}} \sqrt{ I_j \pm \sqrt{I_j^2 - 4
\det{(\tau'_j)}}}
\end{equation}
where
\begin{equation}
I_j := \det{(A_j)} + \det{(B_j)} + 2 \det{(C_j)}.
\end{equation}
Hence, the contribution of the $j$th global mode to the coherent
information reads
\begin{equation}
J_j = g\left(\sqrt{\det(\sigma'_j)} - 1/2\right) - g(\nu_{j,+} -
1/2) - g(\nu_{j,-} - 1/2).
\end{equation}
The constrained maximization of the total coherent information gives
the quantum capacity of the memory channel per channel use:
\begin{equation}
Q = \frac{1}{n} \max \left\{ \sum_{j=1}^n J_j \right\},
\end{equation}
where the maximum is over the parameters $\{ r_j, t_j \}$ under the
energy constraints (\ref{E_const}). The results of numerical
optimization are plotted in figure \ref{Quantum}.

A simple expression can be written in the limit of infinite
correlations $|s| \gg 1$. For $s_j>0$, we can write the following
asymptotic expressions for the symplectic eigenvalues:
\begin{equation}
\nu_{j,+} \simeq
\sqrt{\eta(1-\eta)\left(t_j+\frac{1}{2}\right)\left(T+\frac{1}{2}\right)e^{s_j-r_j}}
+ O(e^{-s_j/2})
\end{equation}
and
\begin{equation}
\nu_{j,-} \simeq \frac{1}{2} + O(e^{-s_j/2}).
\end{equation}
Analogous expressions can be obtained for $s_j < 0$. Taking in
account the asymptotic expression in (\ref{vout}), it follows that
the coherent information vanishes in the limit
$|s_j|\rightarrow\infty$:
\begin{equation}\label{coherent-info}
{J_\infty}_j = \lim_{s\rightarrow\infty} J_j = 0.
\end{equation}
Hence, we can write the following expression in the limit of
infinite correlations:
\begin{eqnarray}
Q_\infty = \left\{
\begin{array}{cc}
0                & \mbox{if $n$ is even,} \\
\frac{\delta}{n} & \mbox{if $n$ is odd.}
\end{array}\right.
\end{eqnarray}
The finite term
\begin{equation}
\delta = g(N') - g\left(\frac{D+N'-N-1}{2}\right) -
g\left(\frac{D-N'+N-1}{2}\right),
\end{equation}
where
\begin{equation}
N' := \eta N + (1-\eta)T
\end{equation}
and
\begin{equation}
D := \sqrt{(N+N'+1)^2-4\eta N(N+1)},
\end{equation}
comes from the contribution of the global mode with $s_j=0$ (see
\cite{holevo}); however, this contribution becomes negligible if $n
\gg 1$.

\begin{figure}
\centering
\includegraphics[height=0.4\textwidth]{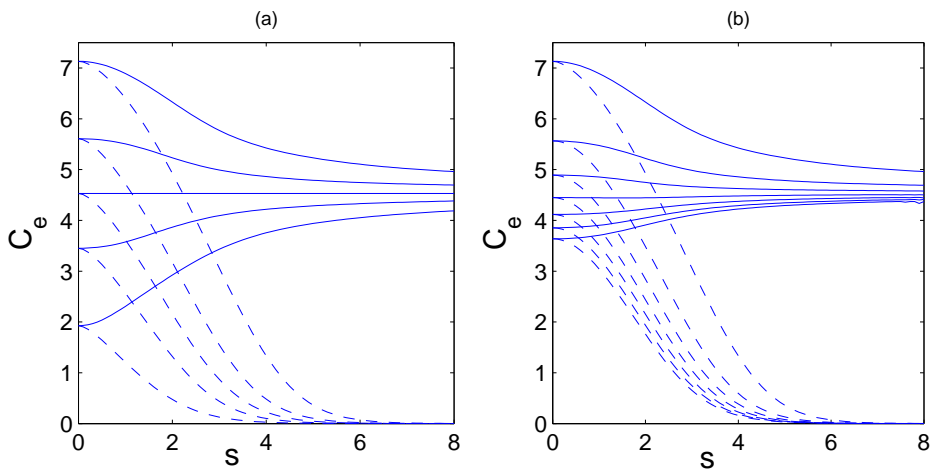}
\caption{The plots show the numerically evaluated
entanglement-assisted classical capacity, for $n=10$, as function of
the memory parameter $|s|$. In (a) at $T=0$ for different values of
$\eta$, from bottom to top $\eta$ varies from $0.1$ to $0.9$ by
steps of $0.2$. In (b) at $\eta=0.9$ for different values of $T$,
from top to bottom $T$ varies from $0$ to $6$ by steps of $1$. The
solid lines refer to the global scenario, the dashed lines to the
local one. The maximum average number of excitations per mode in the
input field is $N=8$.} \label{EntAss}
\end{figure}

The entanglement-assisted classical capacity is obtained maximizing
the quantum mutual information:
\begin{equation}\label{ent_ass}
C_e = \frac{1}{n} \max \left\{ \sum_{j=1}^n I_j \right\}
\end{equation}
where
\begin{equation}
I_j = g(t_j) + J_j.
\end{equation}
The results of numerical maximization are plotted in figure
\ref{EntAss}. In the limit of infinite memory, and for $s_j \neq 0$,
the contribution of the channel $\tilde{\mathfrak{L}}_j^{(1)}$ to
the quantum mutual information is
\begin{equation}
{I_\infty}_j = \lim_{|s_j|\rightarrow\infty} I_j = g(t_j).
\end{equation}
Notice that this asymptotic expression is independent of the
transmissivity $\eta$ and the temperature parameter $T$. Summing
over $j$ we obtain:
\begin{eqnarray}
{C_e}_\infty = \left\{
\begin{array}{cc}
g(N)                    & \mbox{if $n$ is even,} \\
g(N) + \frac{\delta}{n} & \mbox{if $n$ is odd.}
\end{array}\right.
\end{eqnarray}
As for the quantum capacity the term $\delta$ comes from the
contribution of the global mode with $s_j=0$, this contribution
becomes negligible if $n \gg 1$.

Concerning the local scenario, figures \ref{Quantum}, \ref{EntAss}
show in dashed lines the quantum and assisted capacity computed
applying the formulas in \cite{holevo}.

\section{Conclusion and comments}\label{conclude}

We have presented analytical and numerical results for the
capacities of a lossy bosonic Gaussian channel with correlated
noise. To emphasize the role of correlations, we have compared two
different scenarios. The global one allows preparation of states at
the input field which are entangled among different channel uses.

\begin{figure}
\centering
\includegraphics[height=0.2\textwidth]{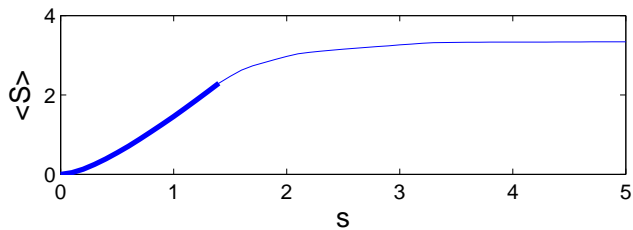}
\caption{The plot shows the von Neumann entropy of the single-mode
reduced state, obtained from the optimal seed state (see
(\ref{seed})), averaged over all the $1:(n-1)$ partitions. This is
for $n=10$, at $T=0$ and $\eta=0.9$. The tick line refers to the
analytical solution. The maximum average number of excitations per
mode in the input field is $N=8$.} \label{Von}
\end{figure}

For our channel model we have shown that the global scenario is
optimal in the presence of memory. In particular, we have shown
that, in a certain range of the parameters, it allows to enhance the
classical capacity over the memoryless (correlation-free) channel.
The optimal seed state of equation (\ref{seed}) turns out to be
entangled as shown by figure \ref{Von} where the von Neumann entropy
of the reduced state, averaged over all the $1:(n-1)$ partitions, is
plotted. The global scenario also allows to enhance the
entanglement-assisted classical capacity, at least for $\eta<0.5$.
Moreover, it slows down the decrement of the quantum capacity, being
the latter a decreasing function of the memory parameter. It is
worth noticing that all the results can be generalized to a broad
class of bosonic Gaussian channels with correlated noise. This
channels are those defined by an environment covariance matrix that
can be diagonalized by a transformation which is symplectic and
orthogonal (see \ref{appxA}).

\begin{figure}
\centering
\includegraphics[height=0.5\textwidth]{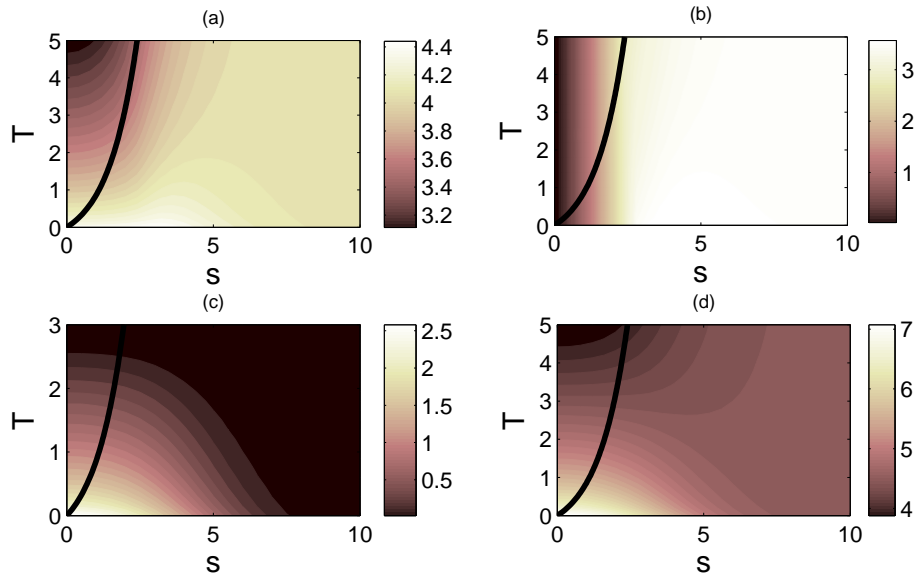}
\caption{For $n=2$, the contour plots show: (a) the lower bound for
the classical capacity, (b) the von Neumann entropy of the reduced
state of the optimal seed state, (c) the quantum capacity and (d)
the entanglement-assisted classical capacity, as function of the
parameters determining the state of the two-mode environment: the
memory parameter and the temperature parameter. The value of the
transmissivity is $\eta=0.9$, the maximum average number of
excitations per mode in the input field is $N=8$. The black line
indicates the boundary between the region in which the state of the
environment is separable (on the left) and entangled (on the
right).} \label{TwoUses}
\end{figure}

Finally, we comment on the role of the environment temperature. From
the analytical and numerical results we can deduce that increasing
the temperature of the environment is qualitatively equivalent to
decreasing the beam splitter transmissivity. Hence, by increasing
the environment temperature more noise is injected in the channel
without any qualitative change in the behavior of its capacities. It
is also worth noticing that, for fixed nonvanishing value of the
squeezing parameter $s$, by increasing the temperature the
environment state makes a transition from entangled to separable.
However, from the point of view of the channel capacities we do not
find any evidence of this transition. In particular, all the
(analytical and numerical) results are smooth functions of the
environment parameters, moreover no qualitative difference is found
in the pattern of the capacities at the transition form classical to
quantum correlations. As an illustrative example we present in
figure \ref{TwoUses} some plots for the case of two channel uses.

\ack

This work has been supported by EU under project CORNER (number
FP7-ICT-213681).

\appendix

\section{Capacities of a broad class of bosonic Gaussian channels with correlated noise}\label{appxA}

In the global scenario the channel $\mathfrak{L}^{(n)}$ is unitary
equivalent to a correlation-free channel which is the product of $n$
single-mode channels. This equivalence was already discussed in
\cite{mancini}, however here the unitary equivalence preserves the
form of the energy constraints. This property belongs to a large
class of Gaussian memory channels. All the qualitative features
regarding capacities are shared by all the channels belonging to
this class. These channels are those defined by an environment
covariance matrix which is diagonalized by an orthogonal
transformation which is also symplectic (in optics this is called
passive transformation). Let us recall that the action of such a
transformation on the phase space coordinates $(Q_1, Q_2, \dots Q_n,
P_1, P_2, \dots P_n)^\mathsf{T}$ is by a matrix of the form (see
e.g.\ \cite{paris})
\begin{eqnarray}
O = \left( \begin{array}{cc}
\mathbb{X} & \mathbb{Y} \\
-\mathbb{Y} & \mathbb{X}
\end{array}\right),
\end{eqnarray}
with
\begin{eqnarray}
\mathbb{X} \mathbb{X}^\mathsf{T} + \mathbb{Y} \mathbb{Y}^\mathsf{T} & = & \mathbb{I}, \\
\mathbb{X} \mathbb{Y}^\mathsf{T} - \mathbb{Y} \mathbb{X}^\mathsf{T}
& = & \mathbb{O}.
\end{eqnarray}
It follows that the global scenario discussed here for the model
defined by the environment covariance matrix in (\ref{covm}) can be
equally introduced for all the covariance matrices of the following
form
\begin{eqnarray}\label{diagonable}
V = \left(\begin{array}{cc}
\mathbb{X} D_Q \mathbb{X}^T + \mathbb{Y} D_P \mathbb{Y}^T & \mathbb{Y} D_P \mathbb{X}^T - \mathbb{X} D_Q \mathbb{Y}^T \\
\mathbb{X} D_P \mathbb{Y}^T - \mathbb{Y} D_Q \mathbb{X}^T &
\mathbb{X} D_P \mathbb{X}^T + \mathbb{Y} D_Q \mathbb{Y}^T
\end{array}\right),
\end{eqnarray}
where the diagonal matrices $D_Q$, $D_P$ satisfy the Heisenberg
principle, namely
\begin{equation}
D_Q D_P \ge \frac{\mathbb{I}}{4}.
\end{equation}
The class of covariance matrices of the form (\ref{diagonable})
contains all the pure state covariance matrices and all the mixed
states which are obtained by applying a squeezing transformation to
a thermal state. For these states the global scenario can be defined
as in section \ref{glocal}; all the results concerning the global
scenario can be straightforwardly extended, including its optimality
to achieve the classical and quantum capacity.

\end{document}